\documentclass[12pt]{article}
\usepackage{latexsym}
\usepackage[cp1250]{inputenc}

\title{On the Superintegrability of TTW model}
\author{Cezary Gonera  \thanks{supported by the grant 506/1037 of University of {\L}\'od\'z } \footnote{e-mail: cgonera@uni.lodz.pl}  \\ 
Department of Theoretical Physics and Computer Science\\
University of {\L}\'od\'z \\
Pomorska 149/153, 90 - 236 {\L}\'od\'z Poland.\\}
\date{}

\begin{document}
\maketitle
\begin{abstract}
The superintegrability of so called Tremblay-Turbiner-Winternitz (TTW) model has been conjectured on the basis of the fact that all its trajectories are closed.
This conjecture has been proven using the method based on solving the partial differential equations for two functions having the same
Poisson bracket with the Hamiltonian.\\
In the present short paper we show that superintegrability of TTW model can be established by using well-known elegant techniques 
of analytical mechanics. Moreover, the resulting expression ( after an appropriate ordering ) can be generalized to the quantum-mechanical
case.
\end{abstract}

\newpage

\section{Introduction}

Recently an interesting integrable two-dimensional model has attracted some attention , both on classical and quantum 
levels \cite{b1}$\div $\cite{b4}. Its Hamiltonian reads

\begin{eqnarray}
H = p^{2}_{r} + \frac { p^{2}_{\varphi }}{r^{2}} + \omega ^{2}r^{2} +
 \frac {\alpha k^{2}}{r^{2}cos^{2}(k\varphi )} + \frac {\beta  k^{2}}{r^{2}sin^{2}(k\varphi )},\nonumber\\   \alpha ,\beta  > 0 \label{w1}
\end{eqnarray}

where without loosing generality one can put $k \geq 0 $. The coordinate space is defined by
the inequalities $ 0 < r <\infty $ , $ 0 < \varphi  < \frac{\pi }{2k} $ (actually one can
consider other sectors $ \frac{n\pi }{2k} < \varphi  < \frac{(n+1)\pi }{2k} $).
The study of classical trajectories generated by the Hamiltonian (\ref{w1}) has been
performed in Ref. \cite{b2}. It has been shown there that all bounded trajectories are closed
for all rational values of k. Moreover, the period of motion is $ T = \frac{\pi }{2\omega }$
so it is the same for all (bounded ) trajectories.

These findings strongly suggest that the Hamiltonian (\ref{w1}) is superintegrable. This has been
shown to be the case by Kalnins et al \cite{b4}. The authors of Ref. \cite{b4} used their own 
method based basically on solving the partial differential equations for two functions having 
the same Poisson bracket with the Hamiltonian so that their difference provides an additional
constant of motion.

The aim of the present short note is to show that the superintegrability of the model
can be established using standard techniques of analytical mechanics \cite{b5},
\cite{b6}. We show that the superintegrability takes place iff k is rational and find the explicit 
form of additional integral.

\section{Superintegrability of TTW model for rational k}

The integrability of the model defined by the Hamiltonian (\ref{w1}) is obvious because it admits separation of variables. 
Two independent commuting integrals of motions are the Hamiltonian itself as well as          

\begin{eqnarray}
X_{k} = p^{2}_{\varphi } +
 \frac {\alpha k^{2}}{cos^{2}(k\varphi )} + \frac {\beta  k^{2}}{sin^{2}(k\varphi )} 
\label{w2}
\end{eqnarray}

In oder to find out whether and when our system is superintegrable we construct first the 
action variables. This is done using standard methods \cite{b5},\cite{b6}. The invariant 
Arnold - Liouville tori are given by the equations

\begin{eqnarray}
 p^{2}_{r} + \omega ^{2}r^{2} + \frac{1}{r^2}(p^{2}_\varphi + 
 \frac {\alpha k^{2}}{cos^{2}(k\varphi )} + \frac {\beta  k^{2}}{sin^{2}(k\varphi )})
= E
 \label{w3}
\end{eqnarray}

\begin{eqnarray}
 p^{2}_\varphi + 
 \frac {\alpha k^{2}}{cos^{2}(k\varphi )} + \frac {\beta  k^{2}}{sin^{2}(k\varphi )}
= A
 \label{w4}
\end{eqnarray}
As explained in Ref.\cite{b2}, for bounded trajectories the following inequalities hold : 
$E^2 - 4\omega ^{2} \geq 0 $, $ A\geq 0$, $(A+(\beta  - \alpha )k^{2})^{2} - 4k^{2}\beta A >0$.

Integrating over two generators of homotopy group of Arnold-Liouville torus one finds
(after some trivial shift of $I_{2}$ variable) the action variables ( cf. also \cite{b7}).

\begin{eqnarray}
 I_{1} = \frac{E}{4\omega } - \frac{\sqrt{A}}{2}
 \label{w5}
\end{eqnarray}

\begin{eqnarray}
 I_{2} = \frac{\sqrt{A}}{2k}
 \label{w6}
\end{eqnarray}

which leads to

\begin{eqnarray}
  H = E= 4\omega (I_{1} +  kI_{2})
 \label{w7}
\end{eqnarray}

Therefore, calling $\psi   _{1,2}$  the corresponding angle variables one obtains the equations of motion

\begin{eqnarray}
 \dot  \psi_{1} = 4\omega  
 \label{w8}
\end{eqnarray}

\begin{eqnarray}
 \dot  \psi_{2} = 4\omega k  
 \label{w9}
\end{eqnarray}

Now, for k irrational the trajectory covers densely the invariant torus. Consequently, it cannot result from the intersection
of the torus with some hypersurface corresponding to the constant value of additional globally defined integral of motion.
The system is integrable but not superintegrable.

Assume now that

\begin{eqnarray}
 k = \frac{m}{n}  
 \label{w10}
\end{eqnarray}

with m,n natural with no common divisor. It follows from eqs. (\ref{w8}) and (\ref{w9}) that $m\psi _{1} - n\psi _{2} $
is an integral of motion which is functionally independent on $I_{1,2}$. Its disadvantage is that it is not single-valued
function on phase space. In order to get a single-valued function one has to follow the standard procedure describe, for example,
in \cite{b5}, i.e. to take some periodic function of $m\psi _{1} - n\psi _{2} $, say $\cos(m\psi _{1} - n\psi _{2}) $
or $ \sin(m\psi _{1} - n\psi _{2}) $. In practice, it is more convenient to consider

\begin{eqnarray}
 e^{i(m\psi _{1} - n\psi _{2})} = (e^{i\psi _{1}})^m (e^{i\psi _{1}})^n  
\label{w11}
\end{eqnarray}

and then to take real or imaginary part.

The angle variables are computed according to the formula \cite{b6} ( see also \cite{b7} ).

\begin{eqnarray} 
\psi _{i} = \frac{\partial S(r,\varphi ; I_{1},I_{2})}{\partial I_{i}}
 \label{w12}
\end{eqnarray}

where S has the same meaning as in Ref.\cite{b2} once E and A are expressed in terms of action 
variables. Now, $\psi _{i}$ can be readily obtained using eqs. (\ref{w5}),(\ref{w6}) and (\ref{w12}) and the results
of Ref.\cite{b2} ( eqs. (15), (19) and (20) therein ).There is no need to write out explicitly the form of angle variables.
One only notes that $\psi _{1}$ consists of one term while $\psi _{2}$ is a sum of three terms. In the combination 
$m\psi _{1} - n\psi _{2} $, $m\psi _{1}$ cancels against the first term in $n\psi _{2}$. The remaining contributions are
computed from eqs.(19) and (20) of Ref.\cite{b2}. The only subtle point is that one has to find  $\exp{(i\psi) }$ knowing
$\sin\psi $ which a priori is not unique. However, we know that $\exp{(i\psi) }$ is  single-valued on the phase space, 
so we expect that $\cos \psi $ is determined from $\sin\psi $ as a single-valued function; in fact, it
appears that the relevant square root $\sqrt{1 - \sin^{2}{\psi }}$ can be taken explicitly.

Finally, writing out in explicit form eq.(\ref{w11}), we find the following additional integral of motion (actually, we simplified 
final expression by multiplying it by an appropriate function of E and A) :

\begin{eqnarray} 
C = \left(\frac{2\sqrt{A} p_{r}}{r} +i\left(E - \frac{2A}{r^{2}}\right)\right)^m
\left(\sqrt{A} p_{\varphi }\sin{2k\varphi}  + i( (\beta  - \alpha )k^{2} + A\cos{2k\varphi} )\right)^n
 \label{w13}
\end{eqnarray}

One can directly check that C is an integral of motion by taking the time derivative of C and using the Hamiltonian equations of motion 
or by computing the Poisson bracket of C with the Hamiltonian (\ref{w1}). This is completely straightforward. 
Taking real or imaginary part of C one obtains real integral of motion functionally independent on A and E. 
Obviously, we produce in this way only one new independent integral. Moreover, it is also straightforward to see that both real and imaginary
parts contain only either even or odd positive powers of $\sqrt{A}$. Therefore, multiplying by $\sqrt{A}$,
if necessary, we obtain integrals polynomial in momenta $p_{r}$ and $p_{\varphi }$. One can do even better.
Putting $ A = 0 $ in eq.(\ref{w13}) one gets 
\begin{eqnarray}
 C( A=0 ) = i^{m+n}E^{m}((\beta  - \alpha )k^{2})^{n}
 \label{w14}
\end{eqnarray}
Therefore, the real (imaginary) part of C consists of integer powers of A for $ m + n$ even (resp. odd).
By subtracting $C(A=0)$ and dividing by A one obtains a polynomial in momenta $p_{r}$ , $p_{\varphi }$ 
of degree $2(m+n-1)$.The above result can be compared with the findings of Ref. \cite{b4}. Our final form of the integral is a polynomial in
the momenta of degree smaller by one. This is because we made one step more. Namely, as it is explained above, the value of $ C(A = 0)$ (which is a function of energy ) has been subtructed which allowed to extract and neglect the A factor ( because it is the integral of motion by itself ).  

The advantage of the formula (\ref{w14}) is that, under an appropriate ordering, C can be converted into its quantum
counterpart see the forthcoming paper\cite{b8}. It reads 
\begin{eqnarray}
 \hat{C} = \{\frac{-\sin(2k\varphi) }{2k}\frac{d}{d\varphi }(\frac{\sqrt{A}}{k} + 1) - \frac{\cos(2k\varphi) }{2}\frac{\sqrt{A}}{k}(\frac{\sqrt{A}}{k} + 1) + \nonumber\\
 \frac{1}{2}( \lambda  - \nu )(\lambda  + \nu  - 1) \}^{n}
 r^{2m}\{\frac{1}{r^{2}}(H + 2(1 + \sqrt{A})(\frac{1}{r} \frac{d}{dr} - \frac{\sqrt{A}}{r^{2}}))\}^{m}
 \label{w15}
\end{eqnarray}
 where $\lambda $ and $\nu $ are defined by: $\alpha  = \lambda (\lambda  - 1)$ and $\beta  = \nu (\nu  - 1)$ and trivial 
numeric factor has been omitted. Again, the fact that $\hat{C}$ is a constant of the motion can be verified by explicit computation
of the commutator $ [ \hat{C}  , H ]$. Moreover, $\hat{C}$ obeys the correspondence principle because, as it has been mentioned above, it is obtained from its classical counterpart by an appropriate ( although by far nontrivial ) ordering.\\
It will be shown in Ref.(\cite{b8}) that $\hat{C}$  (together with its hermitean conjugate) can be 
used to construct the polynomial (in derivatives) integral of motion.

\section{Conclusions}
 
 To sumarize let us discuss the relation of action-angle variables method used above to the one proposed in 
Refs.\cite{b4} and \cite{b9}. The common idea of both  methods is that one is looking 
for the functions on phase space which develop linearly in time; once they are found their
appropriate linear combination is a constant of motion. In fact, for 2D systems admitting separation of variables the method of 
Refs.\cite{b4} and  \cite{b9} provides a superintegral (i.e. a functionally independent additional integral of motion beyond those following from Liouville itegrability) which can also be obtained directly within Hamilton-Jacobi approach. 
However, in general this integral is not well-defined globally on phase - space. 
This becomes crucial in the case of bounded motion (like in TTW model)
which generically is quasiperiodic and, therefore, the additional integrals obtained in the above way are not separating ones.
It is just a very idea of action-angle variables to normalize the new momenta such that the canonically conjugated 
coordinates are angles with $2\pi $ - periodicity. This allows to decide at once whether well-defined superintegrals 
exist globally and to construct them by quadratures. Moreover, within this framework, it is clear that the additional
integrals do not need to be the polynomials in momenta.

\vspace{12pt}
\par
{\large\bf Acknowledgments}

I thank Prof. Piotr Kosi\'nski  for helpful discussion and Prof. Armen Nersessian for bringing ref.[7] to my attention.\\
The present paper the modified version of the preprint arXiv 1010.2915. I am grateful to anonymous referee for remarks which allowed to improve it
substantially.


\begin{thebibliography}{99}
\bibitem{b1}
F.Tremblay, A. Turbiner, P.Winternitz, Journ.Phys. A 42 (2009), 242001\\
\bibitem{b2}
F.Tremblay, A. Turbiner, P.Winternitz, Journ.Phys. A 43 (2010), 015202\\
\bibitem{b3}
C.Quesne, Journ.Phys. A 43 (2010), 082001\\
\bibitem{b4}
E.G.Kalnins, J.M.Kress , W.Miller,Jr., Journ. Phys. A 43 (2010), 265205\\
\bibitem{b5}
L.Landau, E. Lifshitz, Mechanics, Pergamon Press, 1976\\
\bibitem{b6}
V. Arnold, Mathematical methods of Classical Mechanics, Springer, 1978\\
\bibitem{b7}
O.Lechtenfeld, A.Nersessian, V.Yeghikyan, Phys. Lett. A374 ( 2010 ),4647\\
\bibitem{b8}
C.Gonera et al.; "Superintegrability of quantum  TTW model" , to be submitted\\
\bibitem{b9}
E.G.Kalnins, J.M. Kress, W.Miller, Jr., G.Pogosyan, J.Math. Phys. 43, (2002), 3592\\
\end{thebibliography}
\end{document}